\documentstyle[amssymb,aps,multicol]{revtex}

\begin{document}
\draft

\title{Resonant cancellation of off-resonant effects in
a multilevel qubit}

%\vskip 1.5cm
\author{Lin~Tian$^1$, Seth~Lloyd$^2$}
\address{$^1$Department of Physics, Massachusetts Institute of
Technology, Cambridge, MA, 02139}
\address{$^2$d'Arbeloff Laboratory for Information Systems and 
Technology,
Department of Mechanical Engineering, Massachusetts Institute of
Technology, Cambridge, MA, 02139}
\date{\today}
\maketitle

\begin{abstract}
Off-resonant effects are a significant source of error
in quantum computation. This paper presents a 
group theoretic proof that off-resonant transitions to 
the higher levels of a multilevel qubit can be completely
prevented in principle.  This result can be generalized to 
prevent unwanted transitions due to qubit-qubit interactions.
A simple scheme exploiting dynamic pulse control techniques
is presented that can cancel transitions to higher states
to arbitrary accuracy.
\end{abstract}


%PACS number(s): 03.67.Lx, 03.65.Bz, 89.70.+c
%03.67.Lx Quantum computation
%03.65.Bz Foundations, theory of measurement, miscellaneous theories
%(including Aharonov-Bohm effect, Bell
%inequalities, Berry's phase)
%89.70.+c Information science
%42.50.Lc Quantum fluctuations, quantum noise, and quantum jumps
%03.67.-a Quantum information
%03.65.-w Quantum mechanics (see also 05.30 Quantum statistical mechanics)

\begin{multicols}{2}

%% introduction

Successful quantum computation depends on the accurate manipulation of
the quantum states of the qubits\cite{coherence_Lloyd_DiVincenzo}.  
In practice, qubits are subject to many sources of 
quantum errors including thermal fluctuations of 
the environment\cite{environment_decoherence},
qubit-qubit interactions\cite{qubit_interaction},
and intrinsic redundant degrees of freedom
within a qubit such as the quasiparticle
conduction in the superconducting 
qubits\cite{pc_qubit,charge_qubit},
and the effect of the higher levels in many practical
qubit designs\cite{charge_qubit,ion_off_resonance}.  
This paper proposes a dynamic pulse control technique that 
efficiently eliminates unwanted off-resonance transitions.

%% problem

Various schemes to protect the qubit from qubit errors have been 
proposed that can be divided into two categories. The first one is
the quantum error correction technique
\cite{error_correction_shor,error_correction_steane,error_correction_Laflamme,error_correction_Gottesman,error_correction_Calderbank,error_correction_duan} 
where the qubit state is encoded by
redundant qubits. Different errors project 
the qubit-extra-qubit system into different subspaces
that can be determined by measuring the state of the extra qubits.
By applying a transformation according to the measurement, the 
correct qubit state can be restored.  This approach relies
on large numbers of extra qubits to keep the errors from propagating.
The second approach exploits `bang-bang'
control techniques\cite{Viola_Lloyd_Knill} 
where the dynamics of the qubit and the environment 
is manipulated by fast pulses that flip the qubit.
With the influence of the environment averaged out,
the qubit evolves in the error-free subspace.
This method relies on the ability to apply pulses rapidly 
compared with the correlation time of the environment.
This is an open loop control method.

A particularly important form of intrinsic qubit error comes from 
the off-resonant transitions to the
higher levels of a qubit when the qubit is being operated. 
Real qubits are not $S=1/2$ spins that are perfect
two level systems; redundant levels always exist 
that affect the information
content of the qubit.
The additional interaction that is introduced
to achieve qubit operation by
coupling the lowest two states of the qubit 
almost always includes unwanted couplings
between the lowest two states and the higher levels.
When the interaction is applied
with frequency
$\omega=\omega_{2}-\omega_{1}$,
resonant transition occurs between the lowest two states;
meanwhile
off-resonant transitions to the higher states are also switched on.
These transitions deviate the phase and amplitude of the qubit state 
from perfect Rabi oscillation.
Numerical simulations on the superconducting 
persistent current qubit (pc-qubit)\cite{pc_qubit,simulation_pc_qubit}
show that this deviation
can be severe when
the unwanted couplings 
are of the same order as the Rabi frequency.

In this letter, we study the effect of the higher levels
on qubit dynamics during qubit operation 
by a group theoretic approach. We prove
that the errors can be completely avoided by applying
a time varying operation Hamiltonian.
Then we generalize this result
to the qubit-qubit interaction problem which
can be mapped exactly onto the first one.
Using the idea of 
dynamic pulse control\cite{Viola_Lloyd_Knill},
we design a pulse sequence
that cancels the leakage to the higher levels
to arbitrary accuracy with $O(N)$ number of pulses, 
$N$ being the number of higher levels.

The leakage to higher levels has two significant characteristics. 
First, unlike the environmental fluctuations that affect the
qubit only slightly (less than $10^{-4}$) within one operation,
the leakage changes the qubit dynamics on a time scale $1/\omega_0$ that
is much shorter than the qubit operation time (about $1/\omega_{Rabi}$).
Conventional quantum error correction corrects errors that
occur with small probability and is not a suitable strategy to 
cancel the fast off-resonant transitions. 
Neither can we use the bang-bang method to average 
out\cite{average_Hamiltonian}
the leakage simply by manipulating the
lowest two states,
as the pulse induces these unwanted transitions at the same time.
Second, ignoring all interactions with external
variables, the leakage is coherent, although 
the coherent oscillation will collapse since the revival time 
is too long to be observed due to the large number of
transitions of different frequencies\cite{optics_walls}.
As will now be shown, the coherent nature of the leakage
implies that this type of error can be corrected by applying
a control sequence that coherently modifies the qubit dynamics.

%% theory

Consider a $N$ level quantum system with the Hamiltonian ${\cal H}_0$,
the lowest two states of which are chosen as the qubit states 
$|\uparrow\rangle$ and $|\downarrow\rangle$.
The unitary transformations on this $N$ dimensional
Hilbert space form the $N^2$ dimensional compact 
Lie group ${\rm U}(N)$.  Without
other interaction, the trajectory of the qubit follows
the Abelian subgroup $ \{ e^{-i{\cal H}_0t}, t\in R \} $.

Now apply to the qubit the perturbation ${\cal H}_I$,
$[{\cal H}_0, {\cal H}_I]\ne 0$,
to induce a desired transformation of the qubit.
In most physical systems, unwanted
transitions to the higher levels are simultaneously
induced. For example, 
in the pc-qubit\cite{pc_qubit} operation,
$({\cal H_I})_{mn}=2\pi\delta f
\langle m|\sin{(2\phi_m+2\pi f)}|n\rangle\cos{\omega t}$,
when the bias flux is modulated with rf components of
amplitude $\delta f$ and frequency $\omega$.
This perturbation has couplings between  all the levels.
By successive commutation of 
${\cal H}_0$, ${\cal H}_I$, and their commutators
until no independent operator appears,
a Lie algebra ${\cal A}_I$ is created.
In almost all cases, $ {\cal A}_{\rm I} = {\rm u}(N) $\cite{Lloyd_universal},
${\rm u}(N)$ being the Lie algebra of ${\rm U}(N)$.  The only exception occurs
in a zero measure subspace of ${\rm u}(N)$ 
when ${\cal H}_I$ and ${\cal H}_0$ are both in the same
subalgebra of ${\rm u}(N)$. Thus, with almost
all perturbations, the evolution operator
can be any element in ${\rm U}(N)$; and 
transitions to higher levels are unavoidable
with an initial state that only occupies the lowest two levels.

To prevent the transitions to the 
higher states at time $t$ means to restrict the evolution
operator ${\cal U}(t)$ to the submanifold of 
${\rm U}(2)\oplus {\rm U}(N-2)$, ${\rm U}(2)$
being the unitary group on $\{|\uparrow\rangle, |\downarrow\rangle\}$
and ${\rm U}(N-2)$
on the remaining $N-2$ states. 
This applies $4(N-2)$ real domain restrictions on ${\cal U}(t)$:
${\cal U}(t)_{1k}, {\cal U}(t)_{2k} = 0, k=3\, \ldots\,N$.
In contrast to a perfect qubit operation during 
which ${\cal U}(t)$ remains
in the subspace ${\rm U}(2)\oplus {\rm U}(N-2)$ all the time, 
the qubit is allowed to
stray away from this subspace if only it
goes back to this subspace at the designated time $t$. 
The qubit dynamics can be manipulated by varying the
strength and phase of the perturbation with time. 
As the $N^2$ dimensional Lie group ${\rm U}(N)$ is compact, 
any transformation can be reached at time $t$ by adjusting
the $N^2+1$ parameters in the following process\cite{Lloyd_universal}:

 \begin{equation}\label{U_general}
{\cal U}(t)=e^{-i\alpha{\cal H}_It_{N^2}}
e^{-i\alpha{\cal H}_0t_{N^2-1}}\,\cdots\,e^{-i\alpha{\cal H}_0t_1}
 \end{equation}
where $\alpha$ is introduced to ensure that $t=\sum t_i$.
By playing with the $N^2+1$ real parameters,
the $4(N-2)$ real numbers in ${\cal U}(t)_{1k}, {\cal U}(t)_{2k}$
can be set to zero so that the state of the qubit stays
in the $\{\uparrow, \downarrow\}$ space without
leakage. 
Hence by turning the perturbation on and off $O(N^2)$ times,
the lowest two states are completely decoupled
from the higher states. $O(N^2)$ pulses give a sufficient condition
that is required to achieve arbitrary transformation. As will be shown
later in this letter, with proper arrangement,
we can design a pulse sequence of $O(N)$ pulses
to cancel the transitions to the higher levels.
Unlike that in the quantum 
Zeno effect\cite{QZE_measurement}
where measurement is used to prevent the system from evolving,
the dynamics in this process is described
completely by unitary evolutions.

As the unwanted transitions are off-resonant
transitions whose amplitudes decrease roughly as
$\gamma_{ij}/\omega_i$ ($\omega_i$ is the energy of the
$i$th level, $\gamma_{ij}$ the coupling between level $i$ and $j$), 
the influence of the levels with $\omega_i/\omega_0\gg 1$
can be ignored.
In the pc-qubit\cite{pc_qubit}
the energies of the lower levels increase fast enough
($\omega_{10}> 10 \omega_0 $) that levels beyond $|10\rangle$ can
be ignored.  %EJ=200GHz%EC=1/80%alpha=0.75%f=0.495
%%E=0    0.0414    0.1395    0.1965    0.2405    0.2811    0.2875
%%E=0.3831    0.3973    0.4351    0.4554    0.4839
%%\omega_0=0.0414%\omega_10=0.4351>10*\omega_0 %%
The energy of the $i$th level of
the charge state qubit\cite{charge_qubit}
increases as  $i^2$; less levels affect the qubit dynamics
than that in the pc-qubit.
Hence the number $N$ of the higher states
that are involved in the qubit dynamics in real designs
can be reasonably small. As a result, the number of pulses
in the previous analysis is also reasonable.

%% generalization

One question to ask is whether there is 
any fundamental difference
between the errors due to transitions to 
the higher levels
and those due to the fluctuations of the environmental variables.
Putting it in another way: what is the difference between the intra-qubit
coupling in a multilevel qubit and the qubit-external-system coupling?
In the following we will show that
the $N$-level qubit can be mapped into interacting subsystems,
and vice versa.

Let the initial state of a $N$-level qubit
be $|\Psi_0\rangle = \alpha_1^{(0)} |1\rangle + \alpha_2^{(0)} |2\rangle$,
$|1\rangle$ and $|2\rangle$ being the lowest two states.
To map the qubit into two subsystems, we
divide the $N$ states into two subspaces $SP_1$ and
$SP_2$ by adding the vacuum states $|V_1\rangle$ and $|V_2\rangle$ to
the respective subspaces as $SP_1=\{|V_1\rangle, |1\rangle, |2\rangle\}$
and $SP_2=\{|V_2\rangle, |3\rangle,\,\ldots\, , |N\rangle\}$.
Now the $N$ dimensional
Hilbert space of the original qubit is embedded in
the $3(N-1)$ dimensional direct product space $SP_1\otimes SP_2$.
The states in the expanded space are
$|\overline{\Psi}\rangle =
\sum_{i, j} \beta_{i, j} | b_i^{(1)} \rangle | b_j^{(2)} \rangle$,
where $b_i^{(1)}$ and $b_j^{(2)}$ are basis of 
the two subspaces respectively.
The initial state is
$|\overline{\Psi}_0\rangle=(\alpha_1^{(0)}|1\rangle+\alpha_2^{(0)}|2\rangle)
|V_2\rangle$ in the expanded form.
The unitary transformations in this expanded space 
forms the group ${\rm U}(3(N-1))$.

Perturbation introduces coupling between different states.
When mapped to the expanded space, the perturbation $\overline{{\cal H}}_I$
connects states in the $N$ dimensional subspace
spanned by $\{|1\rangle|V_2\rangle,
|2\rangle|V_2\rangle, |V_1\rangle|3\rangle,\,\ldots\,,
|V_1\rangle|N\rangle\}$.
So $\overline{{\cal H}}_I$ and $\overline{{\cal H}}_0$ 
create
$N^2$ dimensional subalgebra ${\rm u}(N)$ in the expanded space.
Under the perturbation, the wave function in the expanded space
can be described as 
$|\overline{\Psi}\rangle=(\alpha_1|1\rangle+\alpha_2|2\rangle)|V_2\rangle
+\sum_{i=3}^N\alpha_i|V_1\rangle|i\rangle$, 
where $\alpha_i$ are time dependent parameters
evolving with the perturbation.

From this analysis, the higher levels 
in the qubit form an effective environment that interferes
{\bf strongly} with the lowest two levels. Interaction
strength
is the major difference between this effective environment and 
a real one\cite{pc_qubit}.
The couplings between $SP_1$ and $SP_2$ are strong and 
comparable to the Rabi
coupling that realizes qubit operation. In contrast, 
the interactions between the environmental oscillators
and the qubit are weak due to the $O(1/\sqrt{V})$ factor
that originates from the normalization of the extented
modes\cite{environment_decoherence}. So
the thermal fluctuations are not enslaved to the qubit dynamics
and can be 
treated classically.
The strong interaction with the higher levels also 
explains why the error due to leakage
occurs at such a short time that a particular strategy is required to
correct the error.
Another thing to mention is that this effective environment
only comes with qubit operation, 
while the real environment affects the qubit all the time.
Hence we worry about the leakage only during qubit operation
and choose to correct the leakage by controlling the operation process.

By reversing the mapping, interacting qubits
can be modeled as one multilevel quantum system.
One example is two interacting qubits with basis
$|b_i^{(1)}\rangle, i=1\,\ldots\, N_1$, and 
$|b_j^{(2)}\rangle, j=1\,\ldots\, N_2$, respectively.
Labeling the state $|b_i^{(1)}\rangle|b_j^{(2)}\rangle$
as $|(i-1)N_2+j\rangle$, we have
the $N_1N_2$ basis for the equivalent multilevel qubit.
In the same way, $n$ two-level qubits 
form a quantum system of $2^n$
levels. Here the number of states grows exponentially with
the number of qubits as the
entanglement between qubits has to be included 
in a single system now\cite{Lloyd_entanglement}. 
Perturbation applied to one of the qubits
can cause unwanted couplings within the
$2^n$ levels,
and induce off-resonant transitions
that affect the qubit performance. Similar to the couplings
of the multilevel qubit,
these  couplings are also strong and cause fast errors.
Taking the pc-qubit as an example, the
interaction between the two qubits is\cite{pc_qubit}:
${\cal H}_2 =
J_z\sigma_z^{(1)}\sigma_z^{(2)}+J_x(\sigma_z^{(1)}\sigma_x^{(2)}
+\sigma_x^{(1)}\sigma_z^{(2)}) $, where
$J_z$ and $J_x$ terms are due to
the inductance coupling between qubit circuits.
When a $\sigma_x^{(1)}$ term is applied to rotate the first qubit,
the second qubit will be involved and qubit dynamics will
be changed.

Although the mapping of the multilevel qubit and the multiqubit system
into each other
is just another way of looking at the same problem, it
shows that the errors due to the qubit-qubit 
interactions\cite{qubit_interaction}
can be treated by the same approach as is used in 
cancelling the interference of the higher levels.
Again we turn to the idea of coherent pulse control that is exploited
in the higher state problem.
Now the number of pulses increases exponentially with the number
of interacting qubits, 
but it doesn't cause a disaster in real designs where
only the nearest neighbour qubit interactions are important 
and $n$ can be made small in the qubit layout geometry.

%% example 

To illustrate the general idea of exploiting dynamic pulse control
to cancel the errors
due to the higher states,
we give an example of pulse sequence that completely
cancels the transitions
to the higher levels with $O(N)$ pulses.
Let us start from a three level system
with eigenvalues $\omega_i, i=1,2,3$. 
The energy 
difference between level $i$ and $j$ is shorthanded as
$\omega_{ij}$.
An interaction ${\cal H}_{I}$
that couples level $i$ and $j$
by $\gamma_{ij}$
is applied to operate the qubit. When the third level is not
present, $\gamma_{12}$ is the Rabi
frequency of the lowest two states.
For simplicity, we ignore the diagonal couplings $\gamma_{ii}$
as $\gamma_{ii}\ll\omega_{i}$.
As will become clear,
the effectiveness of the designed pulse sequence depends on 
the condition $|\gamma_{ij}/\omega_{ij}|\ll 1$ which
is satisfied in most qubits.
%In the pc-qubit\cite{pc_qubit} 
%the energy splitting of the lowest two states
%is $ 4\, {\rm GHz}$ and the third level is $ 10\, {\rm GHz}$
%away from the second level. The $100\, {\rm MHz}$ Rabi frequency
%is two orders smaller than the
%energy splitting. 

The Hamiltonian in the interaction picture is
${\cal H}_{int}=e^{i{\cal H}_0t}{\cal H}_Ie^{-i{\cal H}_0t}
\cos{(\omega t+\phi)}$,
$\omega$ being the pulse frequency.
The wave function
$ \Psi(t) = [ u\, v\, w ]^T $ evolves according to the equation
$ i\frac{\partial \Psi(t)}{\partial t} = 
{\cal H}_{int}\Psi(t)$. When the perturbation is weak,
this equation is integrated order by order as:

 \begin{equation}\label{integral} \begin{array}{lcl}
\Psi(t)&=&\Psi(0) + \int_0^t dt^{\prime} {\cal H}_{int}(t^{\prime})\Psi(0)
\\
& + & \int_0^t dt^{\prime}\int_0^{t^{\prime}}dt^{\prime\prime} 
{\cal H}_{int}(t^{\prime})
{\cal H}_{int}(t^{\prime\prime})\Psi(0) + \cdots \, ,
\\
\end{array}
  \end{equation}
The cosine function is used in the rf pulse 
instead of the single frequency
rotating wave.
In many systems,
no physical correspondence
of the rotating frame exists.
For example, the circuit of the pc-qubit is biased by
$z$ direction magnetic flux and the perturbation
is high frequency modulation of the $z$ flux.
No rotating frame of transverse
field can be defined for the oscillating
flux.

Our strategy to reduce the unwanted transitions
is to divide the qubit operation into short intervals of $t_0$
and attach additional pulses to each operation pulse to
correct errors from this short interval.
The operation pulse 
is in resonant with $\omega_{21}$
of the lowest two states. Besides rotating the qubit between
the level $1$ and $2$, it brings up 
off-resonant transitions
between the third level and these two levels
through the couplings $\gamma_{13}$ and $\gamma_{23}$.
Then the same perturbation is applied in two
other pulses with different 
frequencies, amplitudes and phases as:
$\alpha_{31}{\cal H}_{I}\cos{(\omega_{31}t+\phi_{31})}$ and
$\alpha_{32}{\cal H}_{I}\cos{(\omega_{32}t+\phi_{32})}$, both
for time $t_0$, to cancel the unwanted transitions to the 
third level. 
This three-piece sequence is repeated 
$\tau_{op}/t_0$ times to finish the qubit operation.
The time $t_0$ satisfies 
$1/\omega\ll t_0\ll 1/\gamma_{ij}, i, j =1, 2, 3$ with both 
$1/\omega_{21}t_0 $ and $\gamma_{ij}t_0$ being small parameters of
the same order.
Thus we have two small parameters in this process.
This is crucial for this 
simple pulse sequence to work.

Starting with an initial wave function $\Psi(0)=[u_0\, v_0\, w_0 ]^T$,
$w_0=0$, 
after the $\omega_{21}$ pulse,
the third level has the component:

 \begin{equation}\label{wave1}
\begin{array}{rl}
w = & u_0(\frac{\gamma_{13}^*(e^{-i(\omega_{21}-\omega_{31})t_0}-1)} 
		{\omega_{21}-\omega_{31}}  -
	      \frac{\gamma_{13}^*(e^{i(\omega_{21}+\omega_{31})t_0}-1)}
		{\omega_{21}+\omega_{31}}) \\ [3mm]
 + &     v_0(\frac{\gamma_{23}^*(e^{-i(\omega_{21}-\omega_{32})t_0}-1)}
		{\omega_{21}-\omega_{32}}  -
	      \frac{\gamma_{23}^*(e^{i(\omega_{21}+\omega_{32})t_0}-1)}
                {\omega_{21}+\omega_{32}})   \\ [3mm]
 + &  u_0 \theta_u + v_0 \theta_v \\
\end{array} ,
 \end{equation}
 %\end{multicols}
where
$\theta_u$ and $\theta_v$ are of third order. 
The main components in $w$ are second order terms
that depend on the initial wave function $u_0$ and $v_0$
linearly. With
$t_0$ satisfying $e^{2i\omega_{21}t_0}=1$,
$u$ and $v$ have third order deviations from the
correct two-level rotation. 
The other two pulses are then applied to 
cancel the
$w$ component.
The $\omega_{31}$ pulse induces a resonant transition between level
one and level three to
cancel the $u_0$ term in $w$; the 
$\omega_{32}$ pulse induces a resonant transition between level
two and level three to
cancel the $v_0$ term in $w$.
The amplitudes and phase shifts of these two pulses can be expanded 
in ascending order as:

 \begin{equation}\label{alpha}
\begin{array}{lcl}
\alpha_{31} e^{i\phi_{31}} & = & \alpha_{31}^{(1)}e^{i\phi_{31}^{(1)}} +
			\alpha_{31}^{(2)}e^{i\phi_{31}^{(2)}} + \cdots \\
\alpha_{32} e^{i\phi_{32}} & = & \alpha_{32}^{(1)}e^{i\phi_{32}^{(1)}} +
                        \alpha_{32}^{(2)}e^{i\phi_{32}^{(2)}} + \cdots \\
\end{array} \ ,
 \end{equation}
The first order coefficients cancel the second order terms in $w$ and
modify the higher order terms $\theta_u$ and $\theta_v$ when:

 \begin{equation}\label{alpha1}
\begin{array}{lcl}
\alpha_{31}^{(1)}e^{i\phi_{31}^{(1)}} & = & 
 \frac{e^{-i(\omega_{21}-\omega_{31})t_0} - 1}{i(\omega_{21}-\omega_{31})t_0}
-\frac{e^{i(\omega_{21}+\omega_{31})t_0} - 1}{i(\omega_{21}+\omega_{31})t_0} \\
\alpha_{32}^{(1)}e^{i\phi_{32}^{(1)}} & = &
 \frac{e^{-i(\omega_{21}-\omega_{32})t_0} - 1}{i(\omega_{21}-\omega_{32})t_0}
-\frac{e^{i(\omega_{21}+\omega_{32})t_0} - 1}{i(\omega_{21}+\omega_{32})t_0} \\
\end{array} \ ,
 \end{equation}
It turns out that the nth order terms of $w$ after the
two correction pulses include
linear terms of $\alpha_{31}^{(n-1)}$ and $\alpha_{32}^{(n-1)}$,
and complicated terms that depend on
$\alpha_{3i}^{(k)}e^{i\phi_{3i}^{(k)}}$ ($k=1\,\ldots\, (n-2)$).
So, for any $n$, $\alpha_{31}^{(n-1)}$ and $\alpha_{32}^{(n-1)}$
can be determined by
the lower order components of $\alpha_{31}$ and $\alpha_{32}$
to cancel the $n$th order of $w$. As a result,
the transitions to the third level can be 
completely erased. The parameters $\alpha_{31}$ and $\alpha_{32}$
do not depend on the initial wave function $u_0$ and $v_0$.
This is similar to solving the wave function
in the perturbation theory where
the higher order terms are derived after the lower order ones.

After the $k$th pulse sequences, with $w=0$, 
the wave function of $u$ and $v$ is:

 \begin{equation}\label{wave_uv}
\left [ \begin{array}{l} u_{k+1} \\ v_{k+1} \\ \end{array}
\right ] = \left [ \begin{array}{ cc } 
\cos{\bar{\varphi}} + \bar{s}_u  & -i\sin{\bar{\varphi}} + \bar{t}_u \\
-i\sin{\bar{\varphi}} + \bar{t}_v & \cos{\bar{\varphi}} + \bar{s}_v \\
 \end{array} 
\right ] 
\left [ \begin{array}{l} u_{k} \\ v_{k} \\ \end{array}
\right ] ,
 \end{equation}
where $\bar{\varphi}=\gamma_{12}t_0$ is the phase rotation
of the two-level qubit; the $\bar{s}$ 
and $\bar{t}$ terms are of third order.
As $w=0$, this is a unitary
transformation that
deviates from the original Rabi oscillation by third order corrections.
The matrix can be written as
${\cal U}(t_0)=\exp{(-i(\gamma_{12}\sigma_x
+\delta_0+\sum_i\delta_i\sigma_i)t_0)}$,
where $\delta_i$ are third order small numbers that
can be determined by known parameters and
do not depend on the index $k$. This
is a renormalization of the qubit operation $\gamma_{12}$ with the third
level decoupled.

This correction strategy is easily generalized 
to $N$($N\ge 3$) level system.
By applying rf pulses with frequencies $\omega_{i1}, \omega_{i2}, i=3...N$,
the transitions to the higher levels are completely erased.
Assuming no particular symmetry
between the states, $2(N-2)$ pulses are required in this process.

One may wonder why this simple pulse sequence works so well
to correct the transitions to the higher states.
For $N-2$ higher levels, to decouple these levels 
is to exert $4(N-2)$ real domain restrictions on the
transformation matrix: 
${\cal U}_{1i}, {\cal U}_{2i} =0, i=3...N$.
Our tools are the Hamiltonians ${\cal H}_0$ and ${\cal H}_{I}$
that create the whole ${\rm u}(N)$ algebra by commutation. 
Our pulse sequence 
${\cal U}(t_0)=\Pi_{i,\beta}P(\alpha_{i\beta}, \phi_{i\beta})
e^{-i\int {\cal H}_{I}\cos{\omega_{21}t^{\prime}}dt^{\prime}} $
($i=3\,\ldots\, N$ and $\beta=\uparrow, \downarrow$ ), 
$P(\alpha_{i\beta}, \phi_{i\beta})
=e^{-i\int {\cal H}_{I}\alpha_{i\beta}
\cos{(\omega_{i\beta}t^{\prime}+\phi_{i\beta})}dt^{\prime}}$,
contents $4(N-2)$ free
parameters. By choosing proper pulse sequence, we can achieve
the decoupling with proper pulse
parameters.

%% ending remark

In conclusion, we discussed the errors due to 
unwanted transitions to the higher states of a qubit
during qubit operation. It was shown by a group theoretic 
argument that these errors can be completely prevented
in principle.
Then we generalized the result to the errors due to qubit interactions,
which can also be prevented when
the number of coupled qubits is not large. 
A simple pulse sequence that
modifies the qubit dynamics and cancels off-resonant transitions 
to arbitrary accuracy with $O(N)$ pulses
was proposed to illustrate the general analysis.
Our results showed that the idea of 
dynamic pulse control\cite{Viola_Lloyd_Knill}
also works for the fast quantum errors due to the higher states
of a qubit.
These results suggest that dynamic pulse control,
together with conventional quantum error correction, can function
as a powerful tool for performing accurate quantum computation in the
presence of errors.

%% acknowledgements

This work is supported by ARO grant DAAG55-98-1-0369
and DARPA/ARO under the QUIC program.

$^\dagger$ tianl@mit.edu;  slloyd@mit.edu

\end{multicols}

\end{document}